\pdfoutput=1
\documentclass[10pt,journal]{IEEEtran}
\usepackage{verbatim}
\usepackage{amsmath,bm}
\usepackage{graphicx}
\usepackage{algorithm}
\usepackage{algorithmic}
\usepackage{booktabs}
\usepackage{multicol}
\usepackage{cite}
\usepackage{mathrsfs}
\usepackage{amsfonts,amssymb}
\usepackage{graphicx}
\usepackage{subfigure}
\usepackage[colorlinks,linkcolor=black,anchorcolor=black,citecolor=black]{hyperref}
\ifCLASSINFOpdf
\else
\fi
\begin{document}
%

\title{Distributed Spatio-Temporal Information Based Cooperative 3D Positioning in GNSS-Denied Environments}

\author{Yue~Cao,
        Shaoshi~Yang,~\IEEEmembership{Senior Member,~IEEE},
        Zhiyong~Feng,~\IEEEmembership{Senior Member,~IEEE},
        ~Lihua~Wang,
        and Lajos~Hanzo,~\IEEEmembership{Life Fellow,~IEEE}

\thanks{Copyright (c) 2015 IEEE. Personal use of this material is permitted. However, permission to use this material for any other purposes must be obtained from the IEEE by sending a request to pubs-permissions@ieee.org.}

\thanks{This work was supported in part by Beijing Municipal Natural Science Foundation under Grant L202012, in part by the National Key Research and Development Program under
Grant 2020YFA0711302, in part by the Fundamental Research Funds for the Central Universities under Grant 2020RC05, in part by the Engineering and Physical Sciences Research Council project EP/P003990/1 (COALESCE), and in part by the European Research Council's Advanced Fellow Grant QuantCom (Grant No. 789028). (\textit{Corresponding
author: Shaoshi Yang})}

\thanks{Y. Cao, S. Yang and Z. Feng are with  Key Laboratory of Universal Wireless Communications, Ministry of Education, and  School of Information and Communication Engineering, Beijing University of Posts and Telecommunications, Beijing, 100876, 
China (E-mails: {caoyue, shaoshi.yang, fengzy}@bupt.edu.cn).

L. Wang is with China Academy of Launch Vehicle Technology, Beijing 100076, China.

L. Hanzo is with School of Electronics and Computer Science, University of Southampton, Southampton SO17 1BJ, U.K. (E-mail: lh@ecs.soton.ac.uk).}

}

\markboth{Accepted to appear on IEEE Transactions on Vehicular Technology, Aug. 2022}%
{Shell \MakeLowercase{\textit{et al.}}: Bare Demo of IEEEtran.cls for Computer Society Journals}

\maketitle

\begin{abstract}
A distributed spatio-temporal information based cooperative positioning (STICP) algorithm is proposed for wireless networks that require three-dimensional (3D) coordinates and operate in the global navigation satellite system (GNSS) denied environments. Our algorithm supports any type of ranging measurements that can determine the distance between nodes. We first utilize a finite symmetric sampling based scaled unscented transform (SUT) method for approximating the nonlinear terms of the messages passing on the associated factor graph (FG) with high precision, despite relying on a small number of samples. Then, we propose an enhanced anchor upgrading mechanism to avoid 
any redundant iterations. Our simulation results and analysis show that the proposed STICP has a lower computational complexity than the state-of-the-art belief propagation based localizer, despite achieving an even more competitive positioning performance.
\end{abstract}

\begin{IEEEkeywords}
 3D, cooperative positioning, factor graph, scaled unscented transform (SUT), wireless localization.
\end{IEEEkeywords}

\IEEEpeerreviewmaketitle

\section{Introduction}
\IEEEPARstart{L}{ocation} awareness plays a crucial role in many emerging applications \cite{lv2016beamspace} relying on wireless networks. Distributed cooperative positioning (CP) \cite{Lv2016, xiong2021cooperative} is a promising technique capable of providing location information for wireless networks operating in global navigation satellite system (GNSS) denied environments, such as forests, tunnels, and underground space. This is a challenging and important scenario for both civilian and military applications.

Distributed CP algorithms are usually based on Bayesian estimation, which treats the node positions as random variables and relies on each node to be localized (usually called \textit{agent}) inferring its own position with the aid of both the internal information from its own hardware and external information gleaned from its cooperating nodes via message passing or belief propagation (BP). Non-parametric BP is a popular self-localization method in sensor networks \cite{Ihler2005}, where the BP messages are approximated by a large number of particles, thus imposing excessive computational complexity. For designing a computationally efficient alternative, the authors of \cite{garcia2017cooperative} proposed a parametric BP approach that utilizes a statistical linear regression (SLR) based affine function for approximating the nonlinear measurement functions. Due to the extra computation of SLR at each linearization iteration, the computational complexity remains high. For better modelling a distributed wireless network using its locality structure,  a class of distributed CP algorithms based on the factor graph (FG) framework were proposed in \cite{Wymeersch2009, wu2015distributed, Wang2021}, where the FG facilitates the evaluation of the marginal function of multivariate global functions more efficiently, hence they are more suitable for distributed implementations than traditional BP. However, as a non-parametric techniques, the so-called sum-product algorithm (SPA) over a wireless network (SPAWN)\cite{Wymeersch2009} also suffers from high computational complexity caused by massive sample points and iterative message computations. The authors of \cite{wu2015distributed} employed the first-order Taylor expansion (TE) for replacing nonlinear terms in the BP messages of the parametric SPA, which however degrades the positioning accuracy. In \cite{Wang2021}, a three-dimensional (3D) universal cooperative localizer (3D-UCL) was proposed for vehicular networks, but no temporal information was exploited. Moreover, the existing CP algorithms usually depend on certain types of ranging observations. For example, both the cooperative least square (CLS) algorithm of \cite{Wymeersch2009} and the SPAWN schemes operate with the aid of time-of-arrival (TOA) and received-signal-strength (RSS).

Against the above backdrop, we propose a low-cost high-performance distributed spatio-temporal information based cooperative positioning (STICP) algorithm for 3D wireless networks operating in GNSS-denied environments. Our STICP has the following benefits: 1) it supports any type of ranging measurements that can determine the distance between nodes. Hence, it is more suitable for large-scale heterogeneous wireless networks than the existing CP algorithms. 2) It exhibits better performance despite its lower computational complexity than the 3D-UCL of \cite{Wang2021}, since both spatial and temporal information\cite{Wymeersch2009} are exploited at a much reduced number of samples. 3) It also outperforms the SPA-TE approach of \cite{wu2015distributed}, since we approximate the nonlinear terms of the messages passing on the FG by a method having higher accuracy at similar computational complexity, but at the cost of higher communication overhead. 4) Our STICP has a much lower computational complexity and communication overhead than SPAWN, and supports a wider variety of ranging measurements than SPAWN, albeit at a marginal performance erosion. These benefits are achieved by using the scaled unscented transform (SUT)  \cite{julier2002scaled} for approximating the messages more efficiently and the enhanced anchor upgrading (EAU) for simplifying the iteration process. Both of analysis and simulation results have demonstrated the advantages of our STICP algorithm.

\textcolor{black}{The rest of this paper is organized as follows. Section II introduces the system model and the  formulation of the cooperative positioning problem. The details of the proposed STICP algorithm, and the analysis of its computational complexity and communication overhead, are presented in Section III. Our simulation results and discussions are given in Section IV, and finally the conclusions are drawn in Section V.}

\textcolor{black}{\textit{Notation:} \(\mathrm{E}\{\cdot\}\) and \(\bm{C}_{\left(\cdot\right)}\) represent the expectation and covariance of a random variable, respectively; \(\|\cdot\|\) denotes the Euclidean norm and \(\delta(\cdot)\) represents the Dirac delta function; \(a\):\(b\) denotes the integer-valued vector of \([a, a+1, \cdots, b]\); \(\sqrt{\ }\) represents the Cholesky decomposition; \((\cdot)_{a}\) represents the \(a\)th column of a matrix; and \((\cdot)^\text{T}\) represents the transpose. Finally, we denote by \(p(\cdot)\) the probability density function (PDF).}

\section{System Model and Problem Formulation}
Consider a wireless network composed of \(N\) agents whose positions are unknown and yet to be estimated, as well as \(A\) anchors whose positions are known \textit{a priori} as reference. We assume that the transmission time is slotted, and the agents can move independently from their positions at time slot \(\left(t-1\right)\) to new positions at time slot \(t\). In the 3D space, the position of \textcolor{black}{node} \(i\) at time slot \(t\) is denoted by \(\bm{x}_{i}^{t}=\left[x_{i}^{t}, y_{i}^{t}, z_{i}^{t}\right]^{\text{T}}(i=0, \cdots, N+A-1)\). Denote the set of anchors from which agent \(i\) receives signals during time slot \(t\) by \(\mathbb{A}_{\rightarrow i}^{t}\), and the set of agents from which agent \(i\) receives signals during time slot \(t\) by \(\mathbb{U}_{\rightarrow i}^{t}\).

At time slot \(t\), agent \(i\) is able to obtain both the external measurements from its neighbors, and the internal measurements based on its own hardware. Denote all the internal and external measurements collected by all the agents at time slot \(t\) as the matrix \(\bm{Z}^{t}\). Note that \(\bm{Z}^{t}\) can be broken up into the matrices  \(\bm{Z}_{\text {self}}^{t}\) and \(\bm{Z}_{\text{rel}}^{t}\), where \(\bm{Z}_{\text {self}}^{t}\) consists of all the internal measurements of all the agents, and \(\bm{Z}_{\text{rel}}^{t}\) consists of all the external measurements of all the agents relative to their individual neighbors. \textcolor{black}{Additionally, the dimension of matrix ${\bm Z}^t$, ${\bm Z}_{\textrm{self}}^t$ and ${\bm Z}_{\textrm{rel}}^t$ is \(\gamma n\times \left(N+N_{\textrm{rel}} N\right)\), \(\gamma n\times N\) and \(\gamma n\times N_{\textrm{rel}} N \), respectively, where \(n\) denotes the dimension of the agent position vector (i.e., 2 and 3 for 2D and 3D positioning,  respectively), \(N_{\textrm{rel}}\) denotes the number of neighbor nodes. and $\gamma$ represents the types of measurements considered. More specifically, we have $\gamma = 1$ if only a single type of information is measured (e.g., position); we have $\gamma = 2$ if two types of information are jointly measured (e.g., position and velocity); and we have $\gamma = 3$ if three types of information are jointly measured (e.g., position, velocity and acceleration). In this paper, we only consider the measurement of the 3D position information, hence we have $\gamma = 1$ and $n = 3$. }

The noise-contaminated ranging measurement\footnote{All the links considered are line-of-sight (LOS), while the non-line-of-sight (NLOS)/LOS mixed environment will be considered in our future work.} from node \(j\) to agent \(i\) at time slot \(t\) can be written as:
\begin{equation}
z_{j \rightarrow i}^{t}=d_{ij}^{t}+e_{j \rightarrow i},
\end{equation}
where \(d_{ij}^{t}\) is the Euclidean distance between node \(j\) and agent \(i\) at time slot \(t\), and \(z_{j \rightarrow i}^{t}\) can be measured in a variety of ways, such as TOA, angle-of-arrival (AOA) and RSS, to name but a few. Furthermore, \(e_{j \rightarrow i}\sim\mathcal{N}\left(0, \sigma_{j \rightarrow i}^{2}\right)\) represents the measurement error that obeys the Gaussian distribution with zero-mean and variance \(\sigma_{j \rightarrow i}^{2}\). The goal of agent \(i\) is to estimate its position \(\bm{x}_{i}^{t}\) at time slot \(t\), given only the external and internal measurements up to time slot \(t\), i.e.,
\textcolor{black}{\(p(\bm{x}_{i}^{t} | \bm{z}_{i,\textrm{self}}^{0:t},\bm{z}_{i,\textrm{rel}}^{0:t})\)}.

We assume that agent \(i\) knows the following information: i) the \textit{a priori} distribution \(p\left(\bm{x}_{i}^{0}\right) \sim \mathcal{N}(\mathrm{E}\{\bm{x}_{i}^{0}\}, \bm{C}_{\bm{x}_{i}^{0}})\) at time slot \(0\), where \(\mathrm{E}\{\bm{x}_{i}^{0}\}=[\mathrm{E}\{x_{i}^{0}\}, \mathrm{E}\{y_{i}^{0}\}, \mathrm{E}\{z_{i}^{0}\}]\) and \(\bm{C}_{\bm{x}_{i}^{0}}=\text{diag}(\sigma_{x_{i}^{0}}^{2}, \sigma_{y_{i}^{0}}^{2}, \sigma_{z_{i}^{0}}^{2})\) represent the expectation and the covariance matrix of \(\bm{x}_{i}^{0}\), respectively; ii) the mobility model characterized by the conditional distribution \(p\left(\bm{x}_{i}^{t} | \bm{x}_{i}^{t-1}\right)\) at any time slot \(t\); iii) the internal measurements \textcolor{black}{\(\bm{z}_{i, \text{self}}^{t}\)} at any time slot \(t\) and the corresponding likelihood function \(p\left(\textcolor{black}{\bm{z}_{i, \text{self}}^{t}} | \bm{x}_{i}^{t-1}, \bm{x}_{i}^{t}\right)\) at any time slot \(t\); iv) agent \(i\) gets other information by data delivery over the wireless network.

\section{The Proposed STICP Algorithm}

\subsection{Derivation of the SUT aided STICP algorithm}

We first factorize \(p(\bm{x}_{i}^{0:t} | \bm{Z}^{1:t})\) as:
\begin{equation}
\begin{aligned} p(\bm{x}_{i}^{0:t} | \bm{Z}^{1:t}) \ \propto \ & p\left(\bm{x}_{i}^{0}\right) \prod_{\tau=1}^{t}\left\{p\left(\bm{x}_{i}^{\tau} | \bm{x}_{i}^{\tau-1}\right)\right.\\ &\left.\times p\left(\textcolor{black}{\bm{z}_{i,\text{self}}^{\tau}} | \bm{x}_{i}^{\tau}, \bm{x}_{i}^{\tau-1}\right) p\left(\bm{Z}_{\text{rel}}^{\tau} | \bm{x}_{i}^{\tau}\right)\right\}. \end{aligned}
\end{equation}

\begin{figure}[t]
	\centering\includegraphics[scale=0.27]{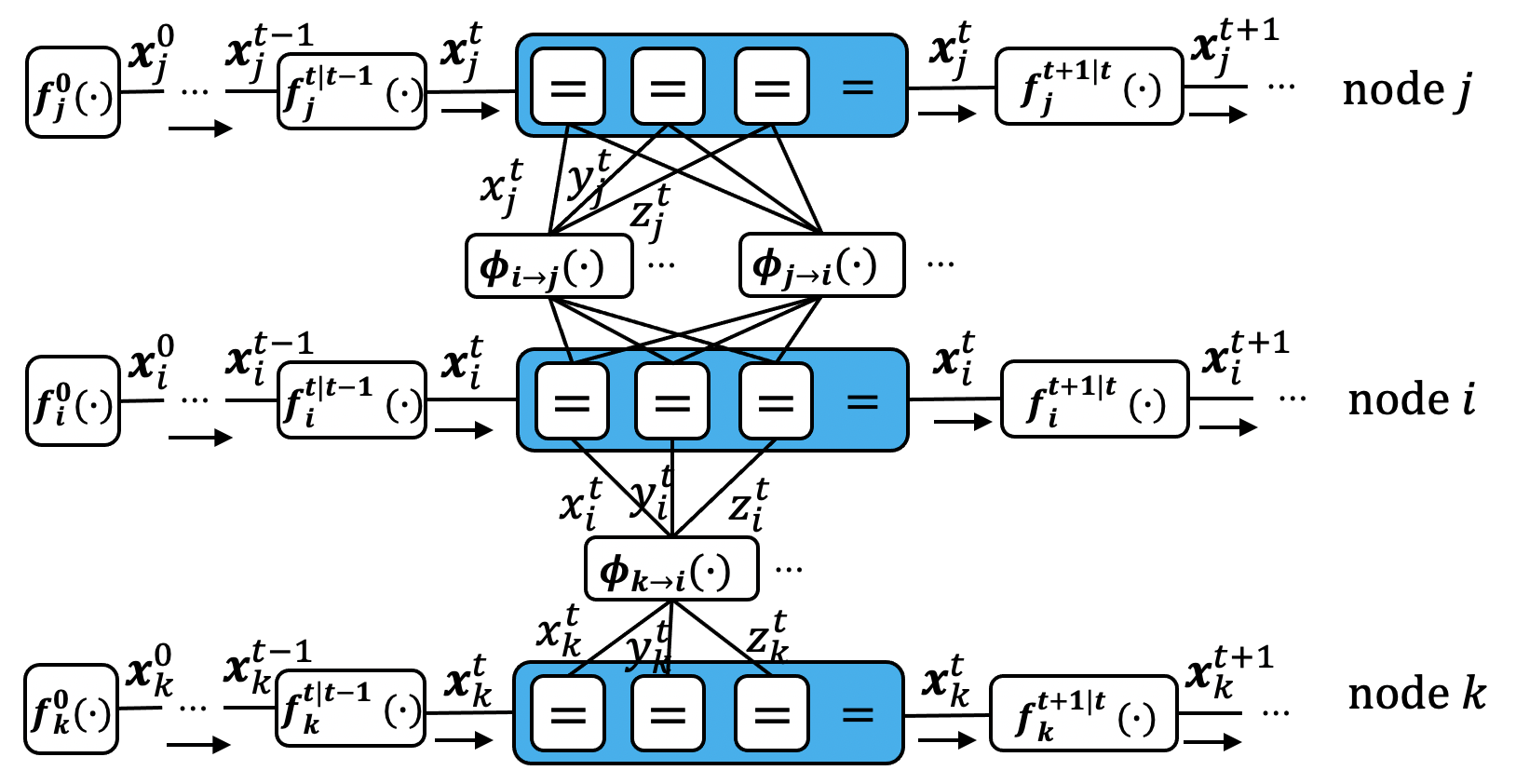}
	\caption{FG of \(p\left(\bm{x}^{0:t} | \bm{Z}^{1:t}\right)\), where nodes \(i, j\in\mathbb{U}_{\rightarrow i}^{t}\), and node \(k\in\mathbb{A}_{\rightarrow i}^{t}\). The arrows represent the temporal flow of the messages passed at different time slots inside a single node (from past to present). We use the following notations: \(f_{i}^{0}(\bm{x}_{i}^{0})=p\left(\bm{x}_{i}^{0}\right)\), \(f_{i}^{t | t-1}(\bm{x}_{i}^{t},\bm{x}_{i}^{t-1})=p\left(\bm{x}_{i}^{t} | \bm{x}_{i}^{t-1}\right) p   \left( \textcolor{black}{ {\bm z}_{i, \text{self}}^{t}} | \bm{x}_{i}^{t-1}, \bm{x}_{i}^{t}\right)\) and \(\phi_{i \rightarrow j}(\bm{x}_{i}^{t}, \bm{x}_{j}^{t})=p\left(z_{i \rightarrow j}^{t} | \bm{x}_{i}^{t}, \bm{x}_{j}^{t}\right)\).}
	\label{fig:1}       
\end{figure}

The Forney-style FG \cite{Forney2001} of \(p(\bm{x}_{i}^{0:t} | \bm{Z}^{1:t})\) has a structure illustrated in Fig.\ref{fig:1}. For each factor, we create a vertex (drawn as a rectangle), and for each variable we create an edge (drawn as a line). When a variable appears in a factor, we connect the edge to the vertex. When a variable appears in more than two factors, an equality vertex is created. For example, the variable \(\bm{x}_{i}^{t}\) appears in factors \(\phi_{i \rightarrow j}(\bm{x}_{i}^{t}, \bm{x}_{j}^{t})\), \(\phi_{j \rightarrow i}(\bm{x}_{i}^{t}, \bm{x}_{j}^{t})\), \(\phi_{k \rightarrow i}(\bm{x}_{i}^{t}, \bm{x}_{k}^{t})\), \(f_{i}^{t | t-1}(\bm{x}_{i}^{t},\bm{x}_{i}^{t-1})\), and \(f_{i}^{t+1 | t}(\bm{x}_{i}^{t},\bm{x}_{i}^{t+1})\), thus we create an equality vertex and label it ``=". Additionally, when agent \(i\) performs external measurement relative to its neighbour node \(j\) at time slot \(t\), we created a factor \(\phi_{j \rightarrow i}(\cdot)\) which is local to agent \(i\) and is a function of the ranging measurement \(z_{j \rightarrow i}^{t}\).
We then run an iterative SPA on the above FG, as detailed below. The \textit{a posteriori} distribution (usually called \textit{belief}) concerning the \(x\)-component of the position vector of agent \(i\) represents the \(x\)-component of the message broadcast by agent \(i\) at iteration \(l\) and time slot \(t\), i.e., \({b}^{l}(x_{i}^{t})\), and it satisfies:
\begin{equation}
\begin{aligned}
{b}^{l}(x_{i}^{t}) \ \propto \ & \mu_{f_{i}^{(t | t-1)}(\cdot) \rightarrow x_{i}^{t}}(\cdot) \prod_{k \in \mathbb{A}_{\rightarrow i}^{t}} \mu_{\phi_{k \rightarrow i}(\cdot) \rightarrow x_{i}^{t}}^{l}(\cdot) \\& \prod_{j \in \mathbb{U}_{\rightarrow i}^{t}}  \mu_{\phi_{j \rightarrow i}(\cdot) \rightarrow x_{i}^{t}}^{l}(\cdot), \end{aligned}
\end{equation}
where \(\mu_{f_{i}^{(t | t-1)}(\cdot)  \rightarrow x_{i}^{t}}(\cdot)\) is the message passed from factor \(f_{i}^{(t | t-1)}(\cdot)\) to variable \(x_{i}^{t}\) and it is the \(x\)-component temporal information of agent \(i\) satisfying:
\begin{equation}
\mu_{f_{i}^{(t | t-1)}(\cdot) \rightarrow x_{i}^{t}}(\cdot) \ \propto \ {b}^{l_\text{max}}(x_{i}^{t-1})f_{i}^{t|t-1}(\cdot),
\end{equation}
with \(l_\text{max}\) representing the maximum number of iterations at time slot \(t-1\), while \(\prod_{j \in \mathbb{U}_{\rightarrow i}^{t}}  \mu_{\phi_{j \rightarrow i}(\cdot) \rightarrow x_{i}^{t}}^{l}(\cdot)\) and \(\prod_{k \in \mathbb{A}_{\rightarrow i}^{t}}\mu_{\phi_{k \rightarrow i}(\cdot) \rightarrow x_{i}^{t}}^{l}(\cdot)\) are the spatial ranging information satisfying:
\begin{equation}
\mu_{\phi_{k \rightarrow i}(\cdot) \rightarrow x_{i}^{t}}^{l}(\cdot) \ \propto \ \iiint \phi_{k \rightarrow i}(\cdot) \mu_{x_{k}^{t} \rightarrow \phi_{k \rightarrow i}(\cdot)}^{l}  (\cdot) d \bm{x}_{k}^{t} d y_{i}^{t} d z_{i}^{t},
\end{equation}
and 
\begin{equation}
\mu_{\phi_{j \rightarrow i}(\cdot) \rightarrow x_{i}^{t}}^{l}(\cdot) \ \propto \ \iiint \phi_{j \rightarrow i}(\cdot) \mu_{x_{j}^{t} \rightarrow \phi_{j \rightarrow i}(\cdot)}^{l}  (\cdot) d \bm{x}_{j}^{t} d y_{i}^{t} d z_{i}^{t},
\end{equation}
respectively. In (5), the message passed from variable \(x_{k}^{t}\) to factor \(\phi_{k \rightarrow i}(\cdot)\), i.e., the belief of node \(k\) broadcast at time slot \(t\) and iteration \(l\) can be written as:
\begin{equation}
\mu_{x_{k}^{t} \rightarrow \phi_{k \rightarrow i}(\cdot)}^{l}(\cdot)=b^{l}(x_{k}^{t})=\delta(x_{k}^{t}-\mathrm{E}\{x_{k}^{t}\}),
\end{equation}
where
\begin{equation}
\begin{aligned}
\begin{array}{c}\phi_{k \rightarrow i}(\cdot) = \frac{1}{\sqrt{2 \pi \sigma_{k \rightarrow i}^{2}}} \exp \left\{-\frac{\left(z_{k \rightarrow i}^{t}-\left\|\bm{x}_{i}^{t}-\bm{x}_{k}^{t}\right\|\right)^{2}}{2 \sigma_{k \rightarrow i}^{2}}\right\}.\end{array}
\end{aligned}
\end{equation}
For brevity, we define the exponential function in (8) as \(h_{k}\), where \(k \in \mathbb{A}_{\rightarrow i}^{t} \cup \mathbb{U}_{\rightarrow i}^{t}\). Then substituting (7) and (8) into (5) leads to:
\begin{equation}
\mu_{\phi_{k \rightarrow i}(\cdot) \rightarrow x_{i}^{t}}^{l}(\cdot) \\ \propto \iiint h_{k} d \bm{x}_{k}^{t}d y_{i}^{t}d z_{i}^{t}.
\end{equation}

Unfortunately, (9) involves integrals and it is difficult to obtain closed-form expressions for (9) due to the term \(h_{k}\), which is a nonlinear function of \(\bm{x}_{i}^{t}\). As a remedy, we approximate the PDF of \(h_{k}\) by employing the SUT technique, \textcolor{black}{which can be used to estimate the result of applying a given nonlinear transformation to the likelihood function of the measurements by picking a minimal set of samples (called \textit{sigma points}) around the mean.} Then the sigma points are used for replacing the \(\bm{x}_{i}^{t}\) in the nonlinear term \(h_{k}\) to form the new mean and variance estimates of \(h_{k}\). In general, it is computationally efficient and convenient to generate a symmetric set of 2\(n\)+1 sigma points to define a discrete distribution having a given mean and covariance in \(n\) dimensions \cite{julier2002scaled}. First, we draw \(N_{s}\)=2\(n\)+1=7 samples \(\{\bm{v}_{a}\}_{a=0}^{6}\) from the PDF of \(\bm{x}_{i}^{t}\), thus we have:
\begin{equation}
\bm{v}_{0}=\mathrm{E}\{\bm{x}_{i}^{t}\}, \quad a=0;
\end{equation}
\begin{equation}
\bm{v}_{a}=\mathrm{E}\{\bm{x}_{i}^{t}\}+\left(\sqrt{(n+\lambda) \bm{C}_{\bm{x}_{i}^{t}}}\right)_{a},
\end{equation}
when \(a=1, 2, 3\);
\begin{equation}
\bm{v}_{a}=\mathrm{E}\{\bm{x}_{i}^{t}\}-\left(\sqrt{(n+\lambda) \bm{C}_{\bm{x}_{i}^{t}}}\right)_{a-3}, 
\end{equation}
when \(a=4, 5, 6\); and
\begin{equation}
\lambda=3 \cdot \left(\alpha^{2}-1\right),
\end{equation}
where \(\alpha\) is a small positive number which controls the distribution of the samples. Then, assign a weight to each sample by:
\begin{equation}
w_{0}^{m}=\frac{\lambda}{n+\lambda},
\end{equation}
\begin{equation}
w_{0}^{c}=\frac{\lambda}{n+\lambda}+\left(1-\alpha^{2}+\beta\right),
\end{equation}
\begin{equation}
w_{a}^{m}=w_{a}^{c}=\frac{1}{2(n+\lambda)}, \quad a=1,2, \ldots, 6,
\end{equation}
where \(w_{a}^{m}\) and \(w_{a}^{c}\) denote the weight assigned to the mean and covariance of the \(a\)th sample, respectively; \(\beta\) is a non-negative number and chosen empirically as 2. 

Similar to \(\bm{x}_{i}^{t}\) in \(h_{k}\), we perform the same operation on \(\{\bm{v}_{a}\}_{a=0}^{2n}\) and obtain \(\bm{g}\)=\(\{g_{a}\}_{a=0}^{6}\):
\begin{equation}
g_{a}=\exp \left\{-\frac{\left(z_{k \rightarrow i}^{t}-\left\|\bm{v}_{a}-\bm{x}_{k}^{t}\right\|\right)^{2}}{2 \sigma_{k \rightarrow i}^{2}}\right\}.
\end{equation}

Then the mean and variance of \(h\) can be obtained upon using the above SUT, and they satisfy:
\begin{equation}
\mathrm{E}\{h_{k}\}=\sum_{a=0}^{6} w_{a}^{m} g_{a},
\end{equation}
\begin{equation}
\sigma_{h_{k}}^{2}=\sum_{a=0}^{6} w_{a}^{c}{(g_{a}-\mathrm{E}\{h_{k}\})}^{2}. 
\end{equation}

Thus the message \(\mu_{\phi_{k \rightarrow i}(\cdot) \rightarrow x_{i}^{t}}^{l}(\cdot)\) is given by:
\begin{equation}
\mu_{\phi_{k \rightarrow i} \rightarrow x_{i}^{t}}^{l}(\cdot) \propto \mathcal{N}(\mathrm{E}\{h_{k}\}, \sigma_{h_{k}}^{2}).
\end{equation}

For brevity, we further denote the message \(\mu_{\phi_{m \rightarrow i}(\cdot) \rightarrow x_{i}^{t}}^{l} (\cdot)\) by \(\mu_{m}^{l}\) (\(m \in \mathbb{A}_{\rightarrow i}^{t} \cup \mathbb{U}_{\rightarrow i}^{t}\)). Moreover, the above derivation also applies to message \(\mu_{j}^{l}\):
\begin{equation}
\mu_{\phi_{j \rightarrow i}(\cdot) \rightarrow x_{i}^{t}}^{l} (\cdot)=\mu_{j}^{l} \propto \mathcal{N}\left(\mathrm{E}\{\mu_{j}^{l}\}, (\sigma_{\mu_{j}^{l}})^{2}\right),
\end{equation}
where
\begin{equation}
\mathrm{E}\{\mu_{j}^{l}\}=(\sigma_{\mu_{j}^{l}})^{2}\left(\frac{\mathrm{E}\{h_{j}\}}{(\sigma_{h_{j}})^{2}}+\frac{\mathrm{E}^{l}\{x_{j}^{t}\}}{(\sigma_{x_{j}^{t}}^{l})^{-2}}\right),
\end{equation}
and
\begin{equation}
(\sigma_{\mu_{j}^{l}})^{2}=\left((\sigma_{h_{j}})^{-2}+(\sigma_{x_{j}^{t}}^{l})^{-2}\right)^{-1}.
\end{equation}

Upon substituting (4), (20) and (21) into (3) we obtain:
\begin{equation}
p(x_{i}^{t}|\bm{Z}^{t})={b}^{l_{max}}(x_{i}^{t}) \propto \mathcal{N}(\mathrm{E}\{x_{i}^{t}|\bm{Z}^{t}\}, \sigma_{x_{i}^{t}|\bm{Z}^{t}}^{2}),
\end{equation}
where
\textcolor{black}{
\begin{equation}
\begin{aligned}
\mathrm{E}\{x_{i}^{t}|\bm{Z}^{t}\}&=\left(\sigma_{x_{i}^{t} \mid \boldsymbol{Z}^{t}}\right)^{2}\left(\frac{\hat{x}_{i}^{t \mid t-1}}{\left(\hat{\sigma}_{i, x}^{t \mid t-1}\right)^{2}}+\sum_{j \in \mathbb{U}_{\rightarrow i}^{t}} \frac{\mathrm{E}\left\{\mu_{j}^{l_{\max }}\right\}}{\left(\sigma_{\mu_{j}^{l_{\text{max}}}}\right)^{2}}\right.\\
&\left.+\sum_{k \in \mathbb{A}_{\rightarrow i}^{t}} \frac{\mathrm{E}\left\{h_{k}\right\}}{\left(\sigma_{h_{k}}\right)^{2}}\right),
\end{aligned}
\end{equation}
}
and
\textcolor{black}{
\begin{equation}
\begin{aligned}
\left(\sigma_{x_{i}^{t} \mid \boldsymbol{Z}^{t}}\right)^{2}&=\left(\frac{1}{\left(\hat{\sigma}_{i, x}^{t \mid t-1}\right)^{2}}+\sum_{j \in \mathbb{U}_{\rightarrow i}^{t}} \frac{1}{\left(\sigma_{\mu_{j}^{l_{\max }}}\right)^{2}}\right.\\
&\left.+\sum_{k \in \mathbb{A}_{\rightarrow i}^{t}} \frac{1}{\left(\sigma_{h_{k}}\right)^{2}}\right)^{-1} \text {. }
\end{aligned}
\end{equation}
}
At any time slot \(t\), each agent can determine the minimum mean squared error (MMSE) based estimate of the \textcolor{black}{\(x\)-component of its own position by taking the mean of \(x_{i}^{t}\), as follows:}
\begin{equation}
\hat{x}_{i}^{t}=\int x_{i}^{t} b^{l_{\text{max}}}\left(x_{i}^{t}\right) \mathrm{d} x_{i}^{t},
\end{equation}
while \(\hat{y}_{i}^{t}\) and \(\hat{z}_{i}^{t}\) can be obtained in a similar manner. Then the resultant STICP is summarized in Algorithm 1.
\begin{algorithm} 
	\caption{STICP} 
	\label{alg3} 
	\begin{algorithmic}
	\footnotesize
		\REQUIRE The \textit{a priori} distribution \(p(\bm{x}_{i}^{0})\), the local mobility model \(p(\bm{x}_{i}^{t}|\bm{x}_{i}^{t-1})\), the local likelihood function \(p(\textcolor{black}{\bm{z}_{i, \text{self}}^{t}} | \bm{x}_{i}^{t-1}, \bm{x}_{i}^{t})\)
		\ENSURE The distribution \(\mathrm{E}\{\bm{x}_{i}^{t}|\bm{Z}^{t}\}\) and \(\bm{C}_{\bm{x}_{i}^{t}|\bm{Z}^{t}}\)
		\FOR{node \(i \in \mathbb{U}\)}
		\STATE initialize \(b^{0}(\bm{x}_{i}^{t})= \mu_{f_{i}^{t|t-1}(\cdot)\rightarrow \bm{x}_{i}^{t}}(\cdot)\).
		\FOR{iteration \(l\) = 1 to \(l_\text{max}\)}
		\STATE broadcast \(b^{l-1}(\bm{x}_{i}^{t})\).
        \STATE receive \(b^{l-1}(\bm{x}_{j}^{t})\).
        \STATE using (3) to calculate the belief.
        \STATE obtain \(\mathrm{E}\{\bm{x}_{i}^{t}|\bm{Z}^{t}\}\) and \(\bm{C}_{\bm{x}_{i}^{t}|\bm{Z}^{t}}\) by (25) and (26).
        \STATE using (27) to estimate the position of node.
		\ENDFOR
		\ENDFOR 
	\end{algorithmic} 
\end{algorithm}

\subsection{Acceleration of STICP}
\textcolor{black}{Based on the change of the estimated position in two consecutive iterations of the proposed STICP algorithm, we propose an EAU technique for further reducing the computational complexity of STICP, by upgrading the agents that meet certain conditions to anchors and by reducing the redundant iterations. The EAU consists of anchor upgrading and iteration reduction. We set two positive thresholds \(\eta_\text{1}\) and \(\eta_\text{2}\), satisfying \(\eta_\text{1}<\eta_\text{2}\), for determining whether the value of the expectation of all components of the position vector, i.e., \(\mathrm{E}\{{\bm x}_{i}^{t}|\bm{Z}^{t}\}\), has met the particular threshold to activate anchor upgrading, or whether the value of the expectation of any components of the position vector, has met the particular threshold to activate iteration reduction.}

\subsubsection{Anchor upgrading}
\textcolor{black}{After several iterations in time slot \(t\),} the individual position estimates of some agents become converged and remain almost unchanged in the subsequent iterations. \textcolor{black}{In particular, if the change of any component of the position vector is less than a given threshold, e.g.,  $\eta_1$,  it is no longer necessary to continue updating the \textit{belief} concerning the particular component of the position vector of agent \(i\) in this time slot. More specifically, if 
\begin{equation}
|\mathrm{E}^{l+1}\{x_{i}^{t}|\bm{Z}^{t}\}-\mathrm{E}^{l}\{x_{i}^{t}|\bm{Z}^{t}\}|<\eta_\text{1}
\end{equation}
is satisfied at time slot \(t\), then we set \({b}^{l+}(x_{i}^{t})={b}^{l}(x_{i}^{t}), l_{+}=l+2,...,l_\text{max}\). Similar operations also apply to $y_i^t$ and $z_i^t$. When the change of all the three components of the position vector, namely when the change of \(\mathrm{E}\{{\bm x}_{i}^{t}|\bm{Z}^{t}\}\), is less than the threshold $\eta_1$ in two consecutive iterations, the agent $i$ is upgraded as a pseudo-anchor.  As a result, the number of agents in the wireless network can be decreased, then it becomes unnecessary to conduct ranging measurements among the pseudo-anchors as well as between a pseudo-anchor and an anchor. Hence the complexity of cooperative positioning is reduced.  }

\subsubsection{Iteration reduction}
\textcolor{black}{In order to further reduce the computational complexity, an iteration reduction technique is invoked. At time slot \(t\), if \(\mathrm{E}\{x_{i}^{t}|\bm{Z}^{t}\}\) changes less than the threshold \(\eta_\text{2}\) but more than the threshold \(\eta_\text{1}\) in two consecutive iterations of STICP, then the update of the \textit{belief} will terminate at the next iteration. Otherwise, the belief is updated normally. More specifically,} if
\begin{equation}
\eta_\text{1}<|\mathrm{E}^{l+1}\{x_{i}^{t}|\bm{Z}^{t}\}-\mathrm{E}^{l}\{x_{i}^{t}|\bm{Z}^{t}\}|<\eta_\text{2},
\end{equation}
then we set \({b}^{l+2}(x_{i}^{t})={b}^{l+1}(x_{i}^{t})\). Therefore, the number of samples is decreased and the computational complexity can be substantially reduced, despite possibly at the cost of marginally increased positioning error, as demonstrated by the subsequent analysis and simulation results. \textcolor{black}{Additionally, \(\eta_\text{1}\) and \(\eta_\text{2}\) are chosen in the range of 0.05$\sim$0.25 and 0.5$\sim$0.8, respectively, based on our experimental experiences. Note that a too large value of either \(\eta_\text{1}\) or \(\eta_\text{2}\) leads to difficulty of convergence, and a too small value leads to unsatisfactory computational speedup.}

\begin{table}[t]
  \scriptsize
  \caption{Comparison of the computational complexity, the communication overhead, and the number of samples (\(N_\text{s}\))}
  \label{tab:1}       
  \centering
   \begin{tabular}{|c|c|c|c|} 
    \hline 
    Algorithm&computational complexity&Communication overhead&\(N_\text{s}\)\\
    \hline  
    STICP&\(\mathcal{O}(N_\text{rel})\)&14&7\\
    \hline
    SPA-TE&\(\mathcal{O}(N_\text{rel})\)&2&N/A\\
    \hline
    3D-UCL&\(\mathcal{O}(N_\text{s} \cdot N_\text{rel})\)&\(\mathcal{O}(N_\text{rel})\)&Large \\
    \hline
    SPAWN&\(\mathcal{O}(N_\text{s}^{2} \cdot N_\text{rel})+\mathcal{O}(N_\text{s})\)&\(\mathcal{O}(N_\text{s})\)&Large\\
    \hline
    \end{tabular}%
\end{table}%
\subsection{Computational Complexity and Communication Overhead}
We compare our STICP, the SPA-TE, 3D-UCL and SPAWN in terms of the computational complexity and the communication overhead in Table~\ref{tab:1}. The communication overhead is evaluated in terms of the number of parameters the agent is expected to broadcast during each iteration. Since the above algorithms are fully distributed, we only consider the agent \(i\). The computational complexity and the communication overhead per SPA-TE \cite{wu2015distributed} iteration are on the order of \(\mathcal{O}(N_\text{rel})\) and \(\mathcal{O}(N_\text{s})\) \cite{wu2015distributed}, respectively, where \(N_\text{rel}\) is the averaged number of cooperative agents. For SPAWN having \(N_\text{s}\) samples, the computational complexity and the communication overhead per iteration are on the order of \(\mathcal{O}(N_\text{s}^{2} \cdot N_\text{rel})+\mathcal{O}(N_\text{s})\) and \(\mathcal{O}(N_\text{s})\) \cite{lien2012comparison}, respectively. With regard to the 3D-UCL, the computational complexity and the communication overhead per iteration are \(\mathcal{O}(N_\text{s} \cdot N_\text{rel})\) and \(\mathcal{O}(N_\text{rel})\), respectively \cite{Wang2021}. For the proposed STICP, all the messages are broadcast and updated as Gaussian distribution, so the number of operations is only related to the number of neighbor nodes \(N_\text{rel}\), which is \(\mathcal{O}(N_\text{rel})\). Furthermore, the communication overhead is quantified as 14 (parameters to be broadcast) since only the approximate mean and the covariance of the seven sigma points have to be broadcast per iteration. \textcolor{black}{Note that the metric ``the number of samples" is not applicable to SPA-TE, because it utilizes the first-order Taylor expansion, instead of a sample-based approach, to approximate the nonlinear term in the messages.} Additionally, the number of samples coming from the SUT step used in STICP is seven, while both SPAWN and 3D-UCL rely on a large number of samples. 

\begin{figure}[t]
	\centering\includegraphics[scale=0.23]{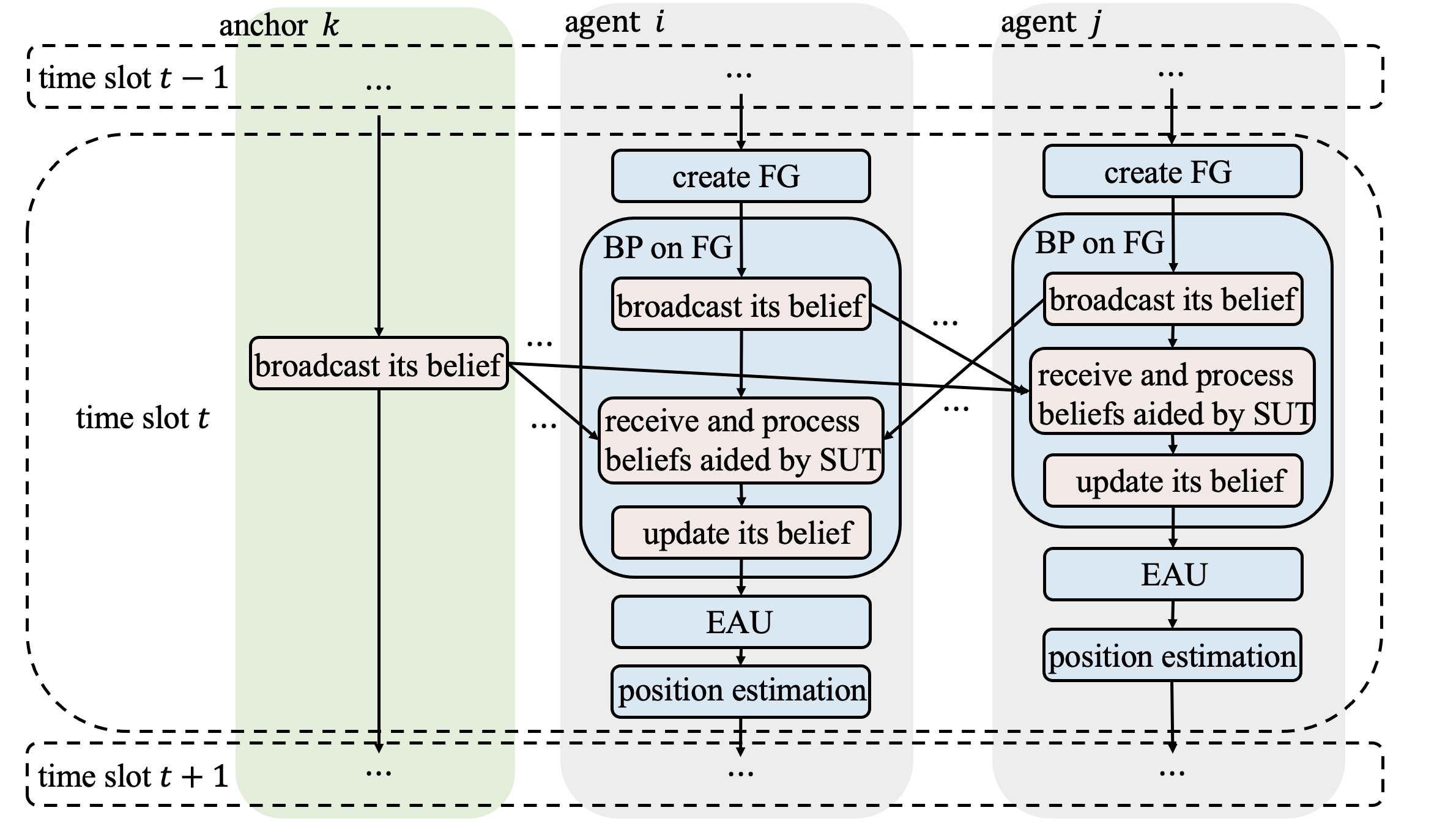}
	\caption{\textcolor{black}{The algorithmic operations conducted on the agent $i$ at time slot $t$.}}
	\label{framework1}       
\end{figure}

\textcolor{black}{For clarity, the logical framework of our STICP is illustrated in Fig. \ref{framework1}. Note that the spatio information, namely the position and ranging information of the spatially distributed nodes, is exchanged between the anchors and the agents, which will affect the traffic in a wireless network \cite{zhong2017heterogeneous}. Meanwhile, the temporal information is the position and ranging information related to any node at different time slots, thus it is local to each node and not exchanged among the nodes. On the other hand, when a distributed wireless network, such as an ad hoc network, is considered, the centralized scheduling is usually infeasible and an random access mechanism is typically used. However, it is also possible to assume that the network has a central scheduler, which may be independent of the anchors and agents, or the anchors may act as the hosts of the scheduler (e.g., multiple base stations can serve as the anchors and host the scheduler). In such a case, we can rely on the contributions of \cite{zhong2017heterogeneous} to analyze the impact of wireless traffic imposed by the cooperative positioning algorithm.}

\section{Simulation Results and Discussions}
We evaluate the performance of our STICP algorithm against several representative CP algorithms by numerical simulations. The positioning performance is quantified by the cumulative distribution function (CDF) \(P(e \leq \epsilon [\text{m}])\), where \(e\) denotes the positioning error and it is characterized by the average mean squared error between the estimated positions and the true position, while \(\epsilon\) represents the tolerable positioning errors. Specifically, we consider a wireless network of 100 agents and 20 anchors, which are randomly scattered across a \((200 \times 200 \times 20) \text{m}^{3}\) space. The distance within which to perform ranging and communication is set to 50\(\text{m}\). The default number of iterations \(l_\text{max}\) is set to 20. Each agent moves a distance \(d_{i}^{t}\) in a random direction, with \(d_{i}^{t}\sim\mathcal{N}(0,1)\). \textcolor{black}{$\eta_1$ and $\eta_2$ are set to 0.18 and 0.6, respectively. The parameter $\alpha$ is set to $3\times 10^{-2}$.}
\begin{figure*}[t]
\centering
\subfigure[TOA measurement]{
\begin{minipage}[t]{0.33\linewidth}
\centering
\includegraphics[width=2.3in]{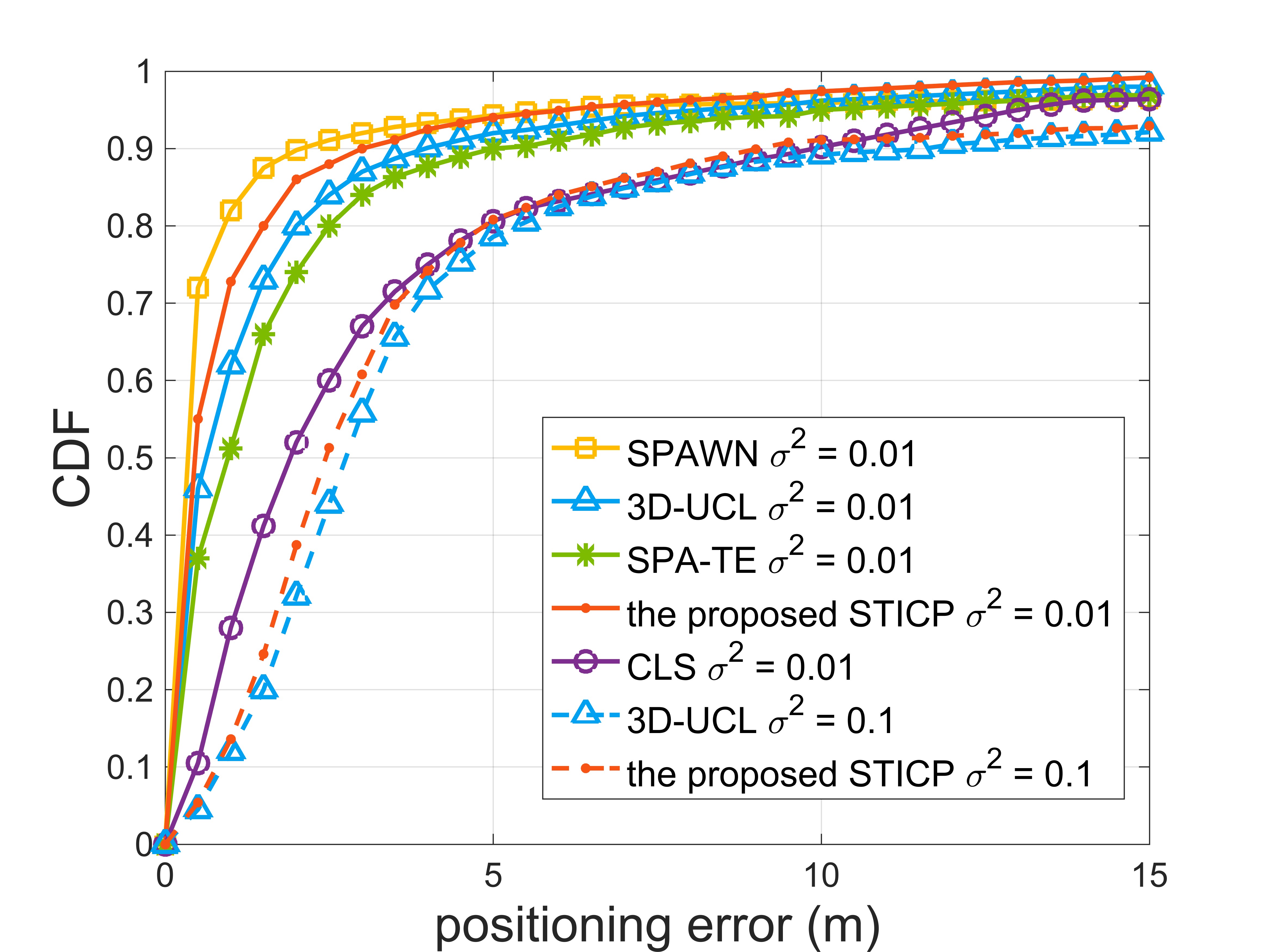}
\end{minipage}%
}%
\subfigure[AOA measurement]{
\begin{minipage}[t]{0.33\linewidth}
\centering
\includegraphics[width=2.3in]{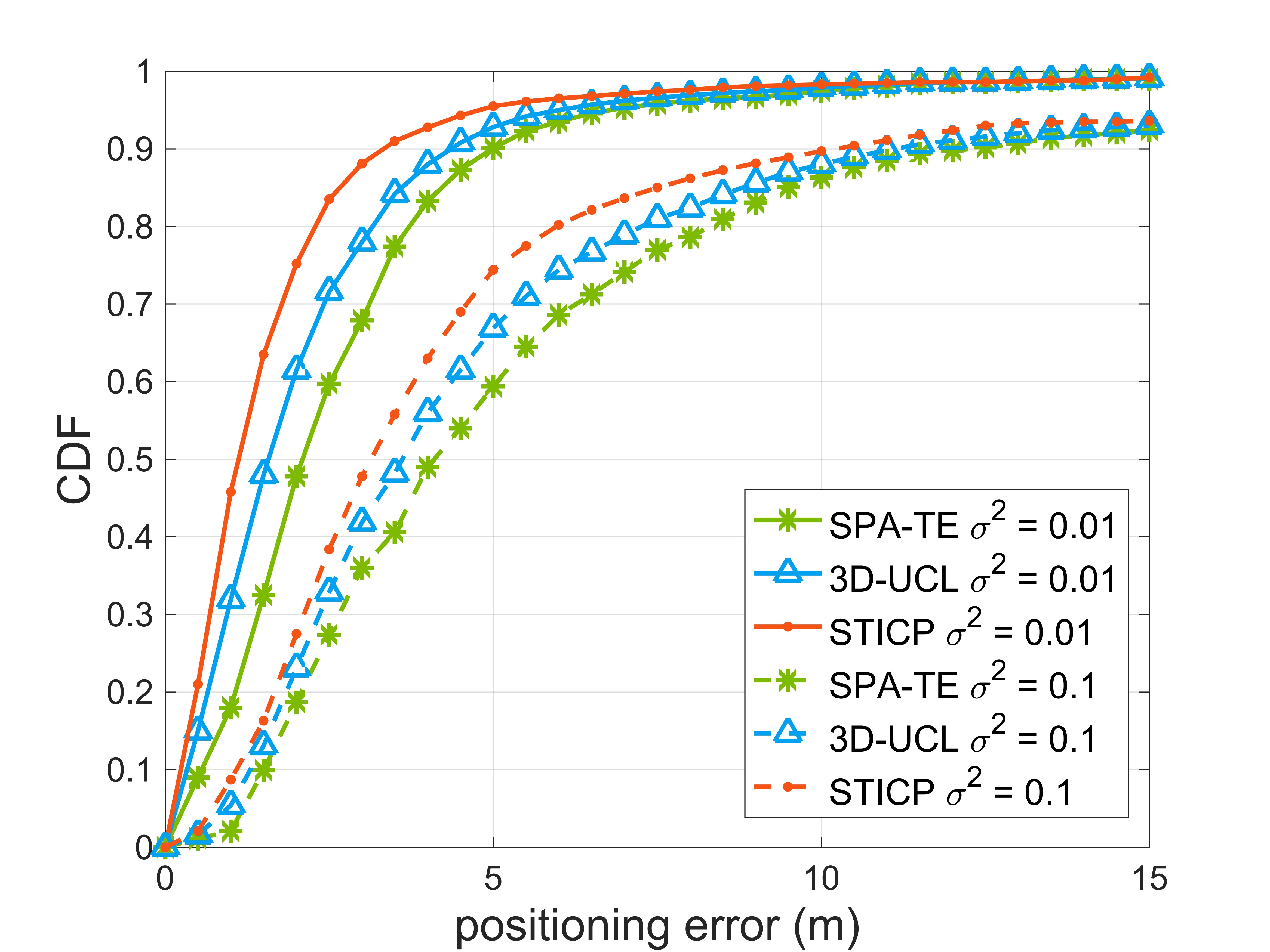}
\end{minipage}%
}%
\subfigure[RSS measurement]{
\begin{minipage}[t]{0.33\linewidth}
\centering
\includegraphics[width=2.3in]{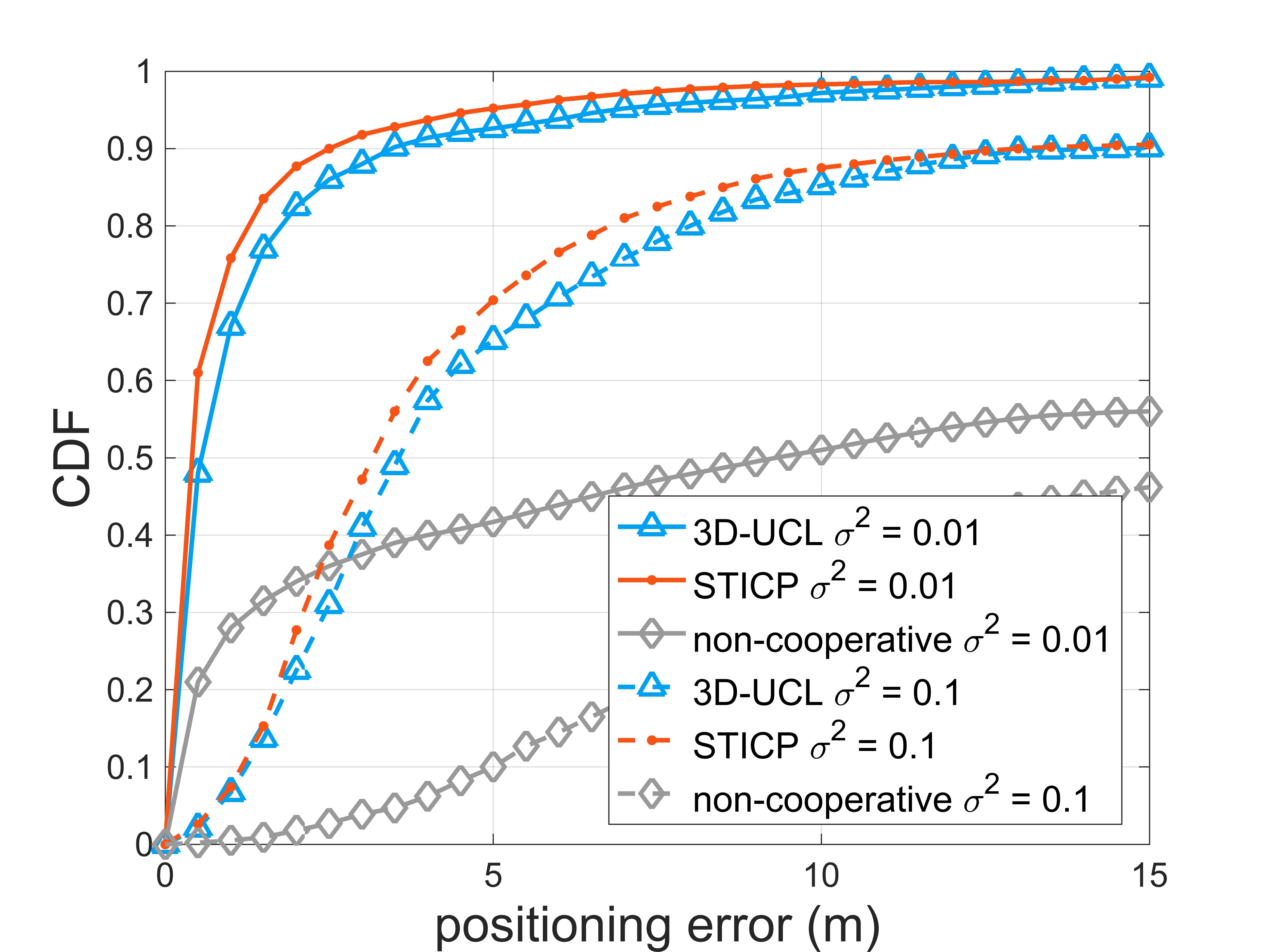}
\end{minipage}
}%
\centering
\caption{CDFs of the positioning error of STICP against representative benchmarking schemes under TOA, AOA and RSS measurements (\(l_\text{max}=20\)).}
\label{fig:2}
\end{figure*}

We first compare the performance of our STICP against the SPA-TE \cite{wu2015distributed}, SPAWN \cite{Wymeersch2009}, 3D-UCL \cite{Wang2021}, and CLS \cite{Wymeersch2009} schemes under TOA measurements in Fig. \ref{fig:2} (a). \textcolor{black}{We observe that STICP outperforms the 3D-UCL, SPA-TE, and CLS schemes when the noise variance is set to \(\sigma^{2}=0.01\).} For example, \(P\left(e \leq 4[\text{m}]\right)\) is 0.93 for STICP, while 0.90, 0.87 and 0.75 for the 3D-UCL, SPA-TE, and CLS schemes, respectively. This is attributed to the gains from internal temporal information. However, STICP performs worse than SPAWN when \(P\left(e\leq2.5[\text{m}]\right)\), due to the use of approximated messages generated by the SUT. But the performance loss is tolerable considering that the computational complexity of STICP is order-of-magnitudes lower than that of SPAWN. \textcolor{black}{Our STICP also outperforms the 3D-UCL under \(\sigma^{2}=0.1\), especially when \(P\left(e\leq4[\text{m}]\right)\)}.

We then compare the performance of our STICP against the SPA-TE and 3D-UCL schemes in Fig. \ref{fig:2} (b) under different noise levels, while relying on AOA measurements. We see that STICP is significantly better than 3D-UCL for \(P\left(e \leq 4[\text{m}]\right)\), and the gain is mainly contributed by the temporal information within agents. Moreover, when \(\sigma^{2}\) is increased from 0.01 to 0.1, the performance degradation of STICP is similar to that of the other benchmarking schemes and the STICP still outperforms the others.

Fig. \ref{fig:2} (c) shows the comparison results of STICP, 3D-UCL and the traditional non-cooperative schemes under different noise levels with RSS measurements. Again, we observe that STICP has the best performance. For instance, when \(\sigma^2 = 0.01\), \(P\left(e\leq 5 [\text{m}]\right)\) is 0.94 for STICP, while 0.93, 0.81, 0.22 and 0.45 for 3D-UCL, CLS and non-cooperative schemes, respectively. When \(\sigma^{2}\) is increased from 0.01 to 0.1, STICP still outperforms the others, which validates that STICP is more robust in noisy environments.
\begin{figure}[t]
	\centering\includegraphics[scale=0.05]{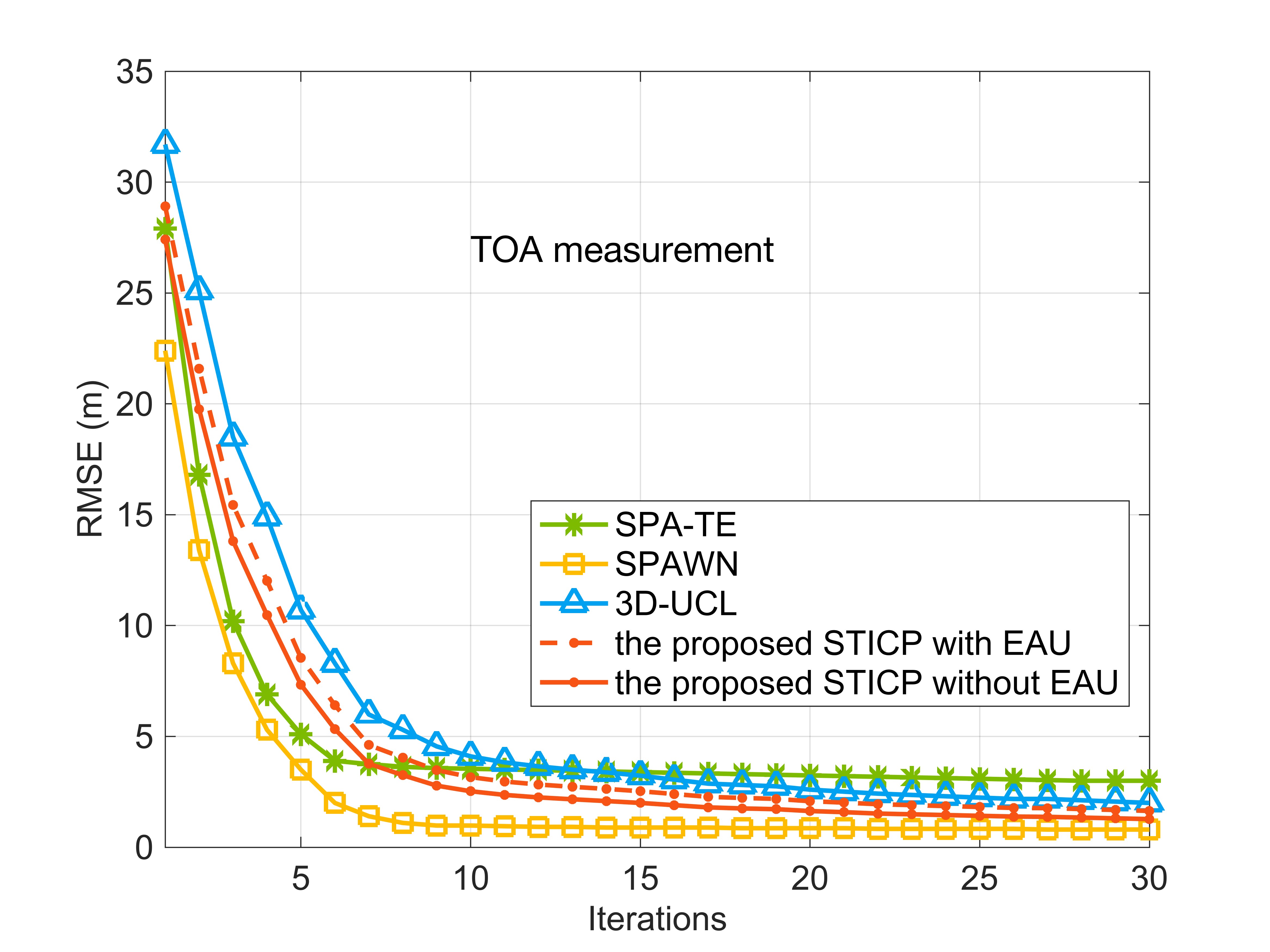}
	\caption{RMSE of the STICP with and without EAU, the SPA-TE, the SPAWN and the 3D-UCL under TOA measurements (noise variance \(\sigma^{2}\) = 0.01).}
	\label{fig:3}       
\end{figure}

The root mean square error (RMSE) performance of the STICP with and without EAU, the SPA-TE, the SPAWN and the 3D-UCL estimators versus the number of iterations under TOA measurements are illustrated in Fig. \ref{fig:3}. Firstly, we can see that the accuracy of all estimators improves upon increasing the number of iterations. Secondly, the RMSE curve of SPA-TE converges quickly around the \(5\)th iteration and SPAWN converges in the \(6\)th iteration. Thirdly, both types of STICP converge faster than 3D-UCL and reach 1.8m and 2.1m positioning accuracy in terms of RMSE after 30 iterations, respectively. The performance degradation of STICP with acceleration may be deemed acceptable in the light of the computational speedup. Last but not least, after 20 iterations, the gains of increasing the number of iterations become insignificant, which allows us to compromise between positioning accuracy and the computational complexity.

\section{Conclusion}

We have proposed a \textcolor{black}{low-complexity high-performance} STICP algorithm for wireless cooperative localization that supports various types of ranging measurements in the GNSS-denied environment. We first create an FG network by factorizing the \textit{a posteriori} distribution of the position-vector estimates and mapping the spatial-domain and temporal-domain operations of nodes onto the FG, \textcolor{black}{which facilitates distributed implementations of wireless cooperative localization.} To approximate the nonlinear terms of the messages passing across the FG, we exploited the symmetric sampling based SUT technique, which achieves high approximation accuracy with a dramatically reduced number of sample points \textcolor{black}{and is able to obtain the closed-form expression of the belief. To further reduce the computational complexity and avoid any redundant iterations, we propose the EAU mechanism to filter out the agents whose position estimates have already converged so that we can terminate the rest of the iterations with minor performance loss.} Our analysis and simulation results validate that the proposed STICP is capable of achieving competitive positioning performances at a low computational complexity.







\bibliographystyle{IEEEtran}
\bibliography{IEEEabrv,reference.bib}


\end{document}